\documentclass[]{llncs}
\usepackage{graphicx}
\usepackage{subfigure}
\usepackage{epsfig}
\usepackage{amsmath}
\usepackage{booktabs}
\usepackage{multirow}

\title{Conditional Restricted Boltzmann Machines for Cold Start Recommendations}

\begin{document}

\author{Jiankou Li\inst{1, 2} \and Wei Zhang\inst{1, 2}}
\institute{
State Key Laboratory of Computer Science, Institute of Software,
Chinese Academy of Sciences, P.O. Box 8718, Beijing, 100190, P.R.China
\and School of Information Science and Engineering,
University of Chinese Academy of Sciences, P.R.China}
\maketitle
\begin{abstract}
Restricted Boltzman Machines (RBMs) have been successfully
 used in recommender systems. However, as with most of other
 collaborative filtering techniques, it cannot solve cold start
 problems for there is no rating for a new item.
In this paper, we first apply conditional RBM (CRBM) which could
take extra information into account and show that CRBM could solve
cold start problem very well, especially for rating prediction task.
CRBM naturally combine the content and collaborative data under
a single framework which could be fitted effectively.
Experiments show that CRBM can be compared favourably with
matrix factorization models, while hidden features learned from
 the former models are more easy to be interpreted.
\end{abstract}

\section{Introduction}
The cold start problem generally means making recommendations
for new items or new users. It has attracted attentions
of many researchers for its importance.
Collaborative filtering(CF) techniques make prediction of unknown 
preference by known preferences of a group of users \cite{su2009survey}.
Preferences here maybe explicit movie rating or implicit item
buying in the online services. Though CF technique has been 
successfully used in recommender system, when there is no rating
 for new items or new users such methods become invalid.
So pure CF could not solve cold start problems.

Different from CF, content-based techniques do not suffer
from the cold start problem, as they make recommendations based
 on content of items, e.g. actors, genres of a movie is usually used. 
However, a pure content-based techniques only recommends 
items that are similar to users' previously consumed items
and so its result lack diversity \cite{park2009pairwise}.

A key step for solving the cold start problem
is how to combine the collaborative information and content
information. Hybrid techniques retain the advantages of CF as
well as not suffer from the cold start problem \cite{agarwal2009regression},
 \cite{gantner2010learning}, \cite{gunawardana2008tied}, 
\cite{jovian2013addressing},  \cite{schein2002methods}.
In the paper, we firstly use condition restricted Boltzmann machines
that could combine the collaborative information and content information
naturally to solve cold start problem.

RBMs are powerful undirected graphical models for rating prediction
\cite{salakhutdinov2007restricted}. Different from directed
graphical models
\cite{agarwal2009regression}, \cite{gantner2010learning},
\cite{gunawardana2008tied}, \cite{jovian2013addressing},  
\cite{schein2002methods}, whose
exact inference is usually hard, inferring latent feature of
RBMs is exact and efficient. An important feature of
undirected graphical model is that they do not suffer the
'explaining-away' problem of directed graphical models
\cite{srivastava2013modeling}.
What's more important is that RBMs could be efficiently
trained using Contrastive Divergence
 \cite{hinton2010practical}, \cite{hinton2002training}.
Based on RBM, CRBM takes the extra information into account
\cite{sutskever2007learning}, \cite{taylor2007modeling}.
The items content feature like actors, genres etc could
be added into CRBM naturally, with little more extra fitting
procedure. 

The remainder of this paper is organized as follows. Section
\ref{related_work} is devoted to related work. Section \ref{the_models}
introduces the models we used including RBM and CRBM. In section
\ref{experiment} we show the results of our experiments.
Finally we conclude the paper with a summary and discuss the
work in the future in section \ref{conclusion}.

\section{Related Work}
\label{related_work}
Cold start problems have attracted lots of researchers.
The key step for solving cold start problem is combining
content feature with collaborative filtering techniques.
On one hand, topic models could model content information
well. On the other hand, matrix factorization is one of the
most successful techniques for collaborative. So most previous works
for solving cold start problem are based on these two techniques.

Previous work \cite{schein2002methods} uses a probabilistic
topic model to solve the cold start problem in
 three step. Firstly, they map users into a latent
 space by features of their related items.
Every user could be represented by some topics of features.
Then a 'folding-in' algorithm is used 
to fold new items into a user-feature aspect model.
Finally, the probability of a user given a new movie
(could be seen as some similarity measure) was calculated in order
to make recommendation.

Gantner et all propose a high-level framework for solving
cold start problem. They train a factorization model 
to learn the latent factors and then learn the mapping
 function from features of entities to their latent
 features \cite{gantner2010learning}.
This kind of model use the content feature while 
retain the advantages of matrix factorization
\cite{koren2009matrix}, \cite{mnih2007probabilistic},
\cite{salakhutdinov2008bayesian} and could generalize well to
other latent factor models and mapping functions.

Another matrix factorization based model is the regression-based
latent factor model (RLFM). 
RLFM simultaneously incorporate user features, item features,
into a single modeling framework.
Different from \cite{gantner2010learning}, where the latent
factors' learning phase and mapping phase are separated,
a Gaussian prior with a feature-base regression was added to the
 latent factors for regularization. RLFM provided a modeling 
framework through hierarchical models which add more 
flexibility to the factorization methods.

In \cite{wang2011collaborative}, the collaborative filtering 
and probabilistic topic modeling were combined through
latent factors. Making recommendations is a process 
of balancing the influence of the content of articles
and the libraries of the other users.
Techniques used for solving cold start problem 
falling into the directed graphical models include
latent Dirichlet allocation \cite{blei2003latent},
\cite{jovian2013addressing},  probabilistic latent
semantic indexing \cite{hofmann1999probabilistic} etc.
Other technique for solving cold start problem include 
semi-supervised \cite{zhang2014addressing}, 
decision trees \cite{sun2013learing} etc.

Besides directed graphical models, undirected graphical models
 also found application in recommender systems.
A tied Boltzmann machine in \cite{gunawardana2008tied}
captures the pairwise interactions between items
by features of items. Then a probability of new items
 could be given to users in order to be ranked.
A more powerful model of undirected graphical model is
Restricted Boltzmann Machines which has found wide application.
RBM was first introduced in \cite{smolensky1986information} named as harmonium.
In \cite{larochelle2008classification}, the Discriminative RBM
was successfully used for character recognition and text
classification. The most important success of RBMs are as initial training
phase for deep neural networks \cite{hinton2007recognize}
and as feature extractors for text and image
\cite{gehler2006rate}.
\cite{salakhutdinov2007restricted} firstly introduce the
RBM for collaborative filtering. In this paper
we will show its good performance for solving cold start problem.

\section{The Models}
\label{the_models}

\subsection{Restricted Boltzmann Machines}

\begin{figure}[h]
\centering
\includegraphics[width=2in]{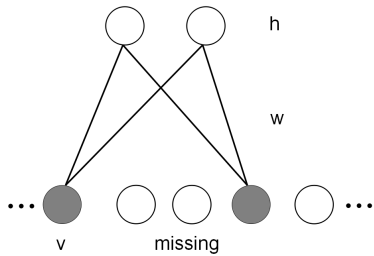}
\caption{Illustration of a restricted Boltzmann machine with
binary hidden units and binary visual units. In this paper,
each RBM represents an movie, each visual unit represents a user that
has rated the item and each hidden unit represents a feature of movies.}
\label{rbm}
\end{figure}

RBM is a two-layer undirected graphical model, see figure \ref{rbm}.
It defines a joint distribution over visible variables \textbf{v} and 
a hidden variable \textbf{h}.
In the cold start situation, we consider the case
where both \textbf{v} and \textbf{h} are binary vectors.

RBM is an energy-based model, whose energy function is given
by 

\begin{equation}
E(v, h) = -\sum_{m=1}^Ma_mv_m - \sum_{n=1}^Nb_nh_n - \sum_{m=1}^M
\sum_{n=1}^Nv_mh_nW_{mn}
\end{equation}

where M is the number of visual units, N is the number of hidden
units, $v_m, h_n$ are the binary states of visible unit
m and hidden unit n, $a_m, b_n$ are their biases and $W_{mn}$
is the weight between $v$ and $h$. Every joint configuration
$(v, h)$ has probability of the form:

\begin{equation}
p(v, h) = \frac{1}{Z}e^{-E(v, h)}
\end{equation}

where $Z$ is the partition function given by summing over
all possible configuration of $(v, h)$:

\begin{equation}
Z = \sum_{v, h}e^{-E(v, h)}
\end{equation}

For a given visible vector $v$, the marginal probability of $p(v)$
is given by
 
\begin{equation}
p(v) = \frac{1}{Z}\sum_he^{-E(v, h)}
\label{marginal_v}
\end{equation}

The condition distribution of a visible unit given all the
hidden units and the condition distribution of a hidden
unit given all the visible units are as follows:

\begin{equation}
p(v_m|h) = \sigma(a_m + \sum_{n=1}^Mh_nW_{mn})
\label{cond_v}
\end{equation}

\begin{equation}
p(h_n|v) = \sigma(b_n + \sum_{m=1}^Nv_mW_{mn})
\end{equation}

where $\sigma(x) = 1/(1 + e^{-x}) $ is the logistic function.

In this paper, each RBM represents a movie, each visual unit
represents a user that has rated the movie. This differs from
\cite{salakhutdinov2007restricted} where each RBM represents a
user and each visual unit represents a movie.
All RBMs have the same number of hidden units and different number
 of visible units because different movies are rated by different users.
 The corresponding weights and biases are tied together in all RBMS. 
So if two movies are rated by the same people, then these two
RBM have the same weights and biases.
In other words, we could say all movies use the same RBM while
the missing value of users that do not rating a specific movie are 
ignored. This idea is similar to the matrix factorization based technique
\cite{koren2008factorization}, \cite{mnih2007probabilistic} that only model
the observing ratings and ignore the missing values.

\subsection{Learning RBM}
We need learn the model by adjusting the weights
and biases to lower the energy of the item while in the same
time arise the energy of other configuration. 
This could be done by performing the gradient ascent in the
log-likelihood of eq. \ref{marginal_v}, 

\begin{equation}
\sum_{t=1}^T\ln p(v^t|\theta)
\end{equation}
where T is the number of movies, $\theta = \{w, a, b\}$ is
the parameter of RBM. The gradient is as follows,

\begin{equation}
\frac{\partial{log p(v)}}{\partial a_{m}} =
\langle v_m \rangle_{data} - \langle v_m \rangle_{model}
\end{equation}

\begin{equation}
\frac{\partial{log p(v)}}{\partial b_{n}} =
\langle h_n \rangle_{data} - \langle h_n \rangle_{model}
\end{equation}

\begin{equation}
\frac{\partial{log p(v)}}{\partial W_{mn}} =
\langle v_mh_n \rangle_{data} - \langle v_mh_n \rangle_{model}
\end{equation}

where  $\langle\rangle$ represents the expectations
under the distribution of data and model respectively.
$\langle\rangle_{data}$ is generally easy to get, while
$\langle\rangle_{model}$ cannot be computed in less than
exponential time. So we use the Contrastive Divergence (CD) algorithm
which is much faster \cite{hinton2002training}.
Instead of $\langle\rangle_{model}$, the CD algorithm
use $\langle\rangle_{recon}$, the 'reconstruction' produced
by setting each $v_m$ to 1 with probability in equation \ref{cond_v}.

Indeed, maximizing the log likelihood of the data is equivalent
to minimizing the Kullback-Liebler divergence between
the data distribution and the model distribution.
CD only minimizing the data distribution and the reconstruction
distribution, and usually the one step reconstruction works well
\cite{salakhutdinov2007restricted}. The update rule is as follows.

\begin{equation}
a_m^{\tau + 1}  = a_m^\tau + \epsilon(\langle v_m \rangle_{data} -
\langle v_m \rangle_{T_1})
\end{equation}

\begin{equation}
b_{n}^{\tau + 1} = a_m^\tau + \epsilon(\langle h_n\rangle_{data} -
\langle h_n \rangle_{T_1})
\end{equation}

\begin{equation}
W_{mn}^{\tau + 1} = W_{mn}^\tau + \epsilon(\langle v_mh_n\rangle_{data} -
\langle v_mh_n \rangle_{T_1})
\end{equation}

where $T_1$ represent the one step reconstruction, $\epsilon$ is the learning
rate.

\subsection{Conditional Restricted Boltzmann Machines (CRBMs)}

The above model could be used in collaborative filtering as 
in \cite{salakhutdinov2007restricted}. But it becomes invalid
for new items as there are no rating and the correspond hidden
units could not be activated. We must add extra
information to this model for solving cold start problem.
Conditional RBM (CRBM) could easily take these extra information
into account \cite{sutskever2007learning}, \cite{taylor2007modeling}.

\begin{figure}[h]
\centering
\includegraphics[width=2in]{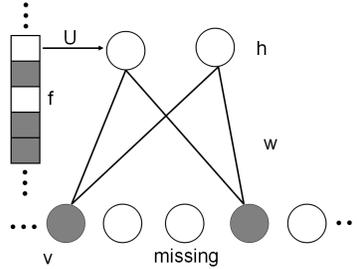}
\caption{Illustration of a conditional restricted Boltzmann machine with
binary hidden units and binary visual units}
\label{cond_rbm}
\end{figure}

For each movie we use a binary vector $f$ to denote its feature 
and then add directed connection from $f$ to its hidden
factors $h$. So the joint distribution over $(v, h)$ conditional
on $f$ is defined, see figure \ref{cond_rbm}.
The energy function becomes the following form.

\begin{equation}
\begin{aligned}
E(v, h, f) = -\sum_{m=1}^Ma_mv_m - \sum_{n=1}^Nb_nh_n -
 \sum_{m=1}^M
\sum_{n=1}^Nv_mh_nW_{mn} -
 \sum_{k=1}^K\sum_{n=1}^Nf_kh_nU_{kn}
\end{aligned}
\end{equation}

The conditional distinction of $v_m$ given hidden 
units is just the same as eq.\ref{cond_v}.
The conditional distribution of $h_n$ becomes
the following form.

\begin{equation}
\begin{aligned}
E(h_n|f, v) = 
\sigma(b_n + \sum_{m=1}^Mv_mw_{mn} + \sum_{k=1}^Kf_kU_{kn})
\end{aligned}
\end{equation}

In CRBM, the hidden states are affected by the feature vector $f$.
When we have new items, there are no rating by users, but
the hidden state could still be activated by $f$, then
the visible unit could be reconstructed as the usual RBM.
What's more, this does not add much extra computation.

Learning the weight matrix U is just like learning the bias
of hidden unit.
In our experiments, we add an regularization term to the normal
 gradient
known as weight-decay \cite{hinton2010practical}. We used the
'$L_2$' penalty function to avoid large weights.
The update rule becomes follows.

\begin{equation}
\begin{aligned}
U_{kn}^{\tau + 1} = U_{kn}^\tau + \epsilon ((\langle h_n\rangle_{data}
- \langle h_n\rangle_{T_1})f_k + \lambda U_{kn} )
\end{aligned}
\end{equation}

Weight-decay for U brings some advantages. On one hand,
it improves the performance of CRBM by avoid overfitting
in the train process. On the other hand, small weight
can be more interpretable than very large values which
will be shown in section \ref{experiment}. More reasons
for weight-decay could be found in \cite{hinton2010practical}.

\section{Experiment}
\label{experiment}
\subsection{DataSet}
We illustrate the results on benchmark datasets (MovieLens-100K).
In this dataset, there are 100K ratings between 943 users
and 1,682 movies. As we mainly focus on the new item cold
start problem, we use the items features only (actors and 
genres). We randomly split the items into two
sets. In the test set there are 333 movies which do not
appear in the train set to used for cold start.
The dataset set comes from the GroupLens project and the
items' feature are downloaded from the
Internet Movie Database (http://www.imdb.com).

\subsection{Recommendation Task}
\label{recommendation_task}
In this section, we make a distinguish among different recommendation
task in real world applications. 
There are mainly three recommendation tasks depending on 
the data we have. They are 
one-class explicit prediction, one-class implicit
prediction, and rating prediction.
%Explicit rating is a rating indicating the user
%likes an item, e.g. a rating $\ge$ 4 in usual rating data. 
%Implicit rating means some feedback of a user which does not
%give explicit preference, e.g. the purchase history, click-through rate etc.

\begin{enumerate}
\item One-class explicit prediction
\\This task is usually arised where we have data
 that only contains explict possitive feedback. Our goal
is to predict whether a user likes an item or not.
In this paper, we convert the original integer rating
 values from 1 to 5 into binary state 0 and 1. 
Concretely, if a rating is bigger than 3,
 we take it as 1, otherwise we take it as 0. As a result,
all the rating not bigger than 3 or missing rating are
 taken to be 0.

\item One-class implicit prediction
\\Implicit rating means some feedback of a user
 which does not give explicit preference, e.g.
 the purchase history, watching habits and browsing activity
 etc. In this task we would predict probability that a user
will rate a given movie. 
In this paper, all the observed ratings are taken to be 1
and all the missing rating are taken to be 0.

\item Rating prediction 
\\Rating imputation is to predict whether a user will like 
an item or not condition on his implicit rating. 
Usually this could be used to evaluate 
an algorithm's performance by holding out some rating
from training data and then predicting their ratings.
In the train phase, all missing value are ignored and
only the observed value are used.
In the test phase, only the observed value in the test
set are evaluated. This is very different from the former
two taskes where all missing value are taken to be 0.
%The visual units represent the 'visual' rating between
%users and items. 1 indicate the user like the item and 0 otherwise.
%The hidden untis represent a latent feature of an item.
\end{enumerate}

In the first two recommendation taskes all the missing
 values are taken to be 
zeros. As a result, the rating matrix becomes dense.
While in the rating prediction task, rating matrix
could remain sparse.
\cite{schein2002methods} also give three recommendation
task which are similar to ours.
It's important to make such a distinction as models can
give notable different performance in difference task. 
In this paper we compare algorithms
on these tasks and analysis their performance.

\subsection{Evaluation Metrics}
Precision-recall curve and root mean square error (RMSE) are two popular
evaluation metrics.
But both of them may be pitfall when the data class has 
skew distribution.
As we know, the rating data is usually sparse and each
user only rates a very small fraction of all the items.
A trivial idea is that we take all the pair as 0, then we will get a not very 
bad result according RMSE or precision-recall curve.
Besides they also
suffer from the problem of zero rating's uncertain.
Zero rating's uncertain here means that a missing rating
may either indicate the user does not like the item or 
that the user does not know the item at all.
%To solve this problem, recall@M curve is used recently  
%\cite{jovian2013addressing},  \cite{wang2011collaborative}.

In this paper, we use receiver operating characteristics (ROC) graph
to evaluate the performance of different models.
ROC graphs have long been used for organizing classifiers and
 visualizing their performance.
One attractive property of ROC curve is that they are
 insensitive to changes in class distribution.
We could also reduce ROC performance to a single scalar value
to compare classifiers, that is the area 
under the ROC curve (AUC) \cite{fawcett2006introduction}.

\subsection{Baselines}

In this paper, we mainly focus on the cold start problem.
As there are so many techniques for recommender systems,
we select three typical models as our baselines.
They are aspect
 model  (AS)\cite{schein2002methods}, tied Boltzmann
machine (TBM)\cite{gunawardana2008tied} and regression-based
latent factor model (RLFM) \cite{agarwal2009regression}.
We select these three models because they
belong to discrete latent factor model, continuous latent
 factor model and undirected graphical models respectively.
They represent three directions for sovling cold start problems.

The former two models both of which are latent factor models
belong to the directed graphical models.
Their difference mainly lie in the type of latent factors.
In the aspect model, the latent
factor is discrete, so aspect model is a mixture model
indeed. RLFM is strongly correlated
with SVD style matrix factorization methods where the latent
factors are continuous. TBM is another type of model belonging
 to the undirected
graphical model which directly models the relationship
between items.

%\subsection{Experiment Settings}
\subsection{Result}

\begin{figure}
\centering 
\subfigure[implicit rating prediction]
{\label{task1_groc}
\includegraphics[width=1.5in]{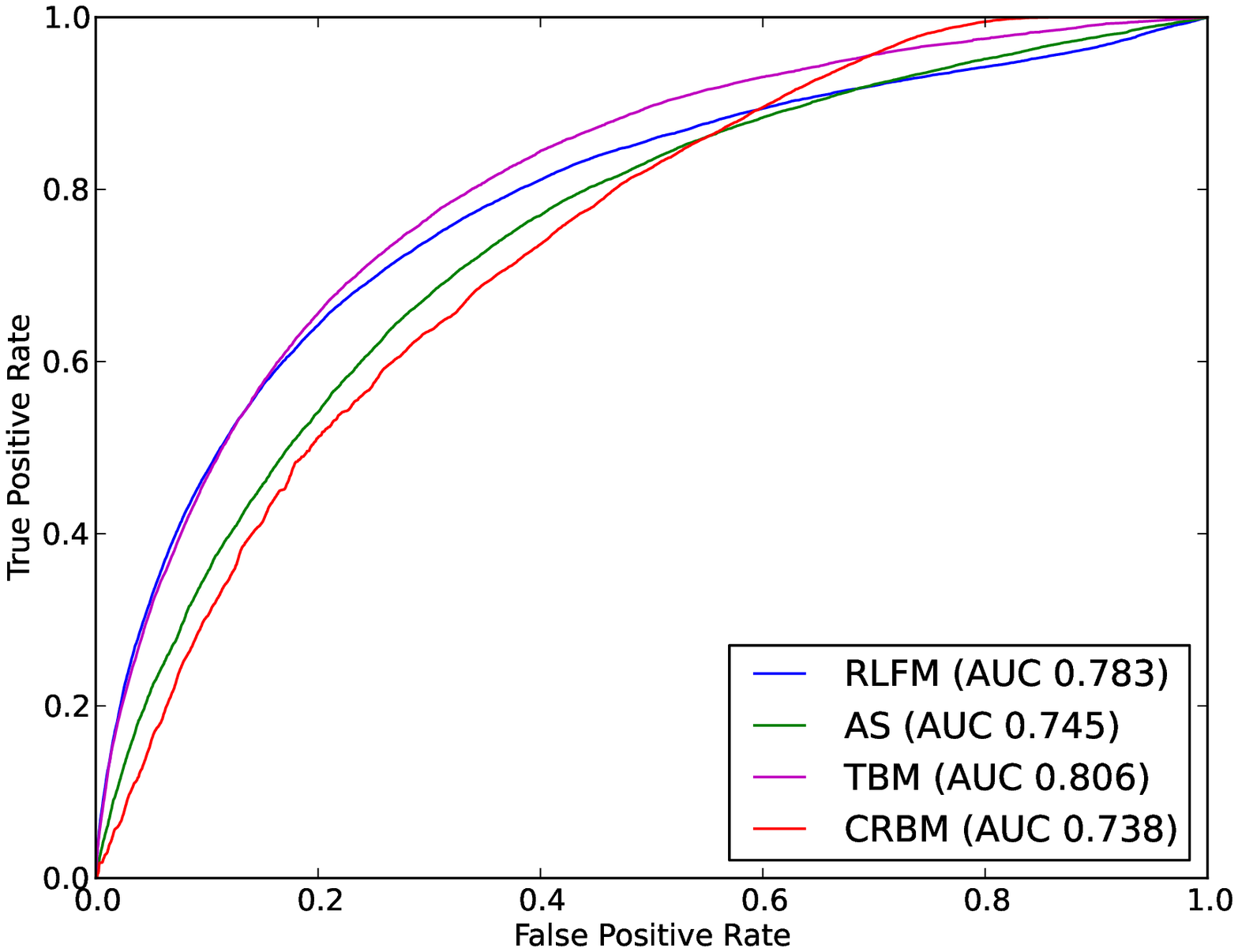}}%
\subfigure[rating prediction]{\label{task2_groc}
\includegraphics[width=1.5in]{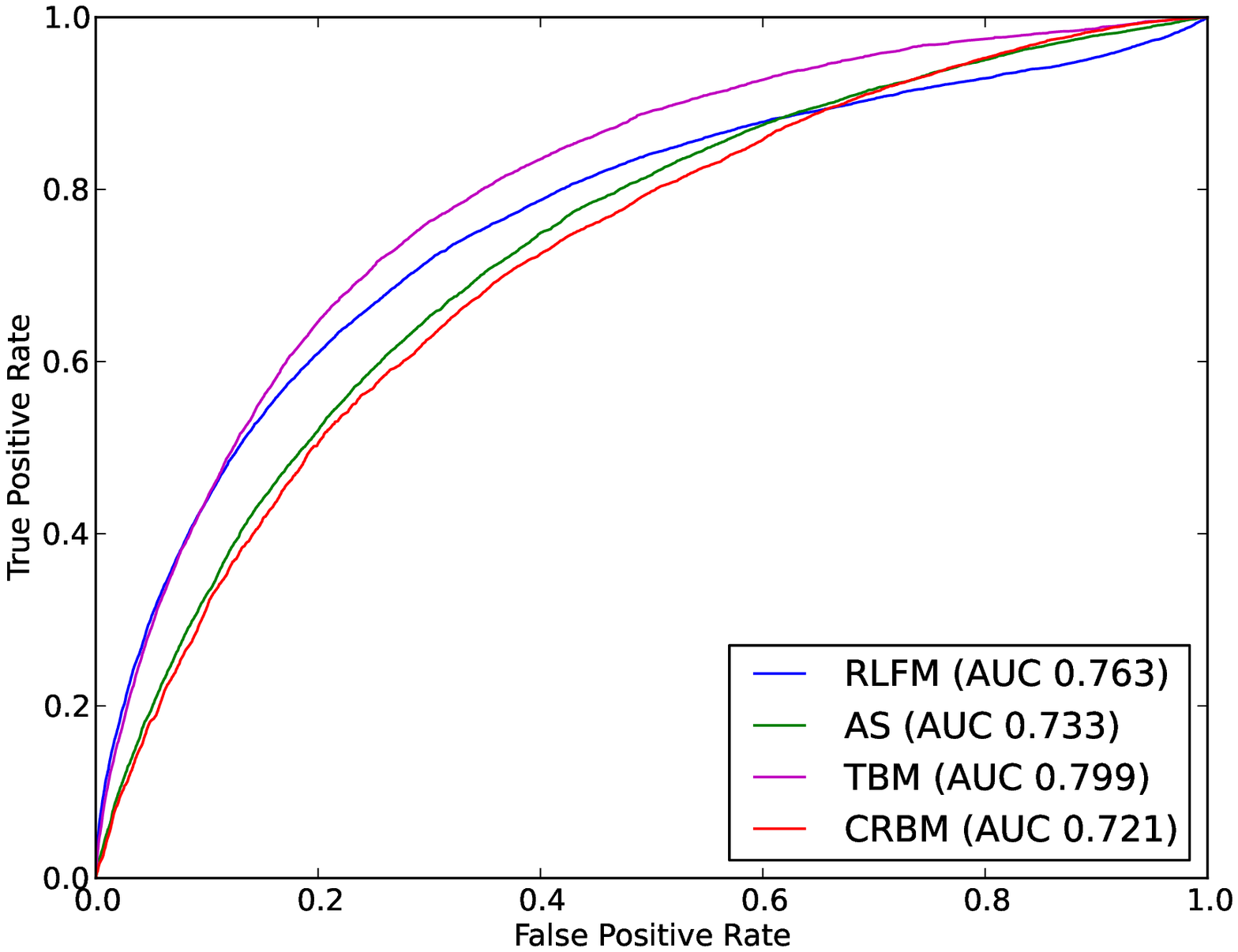}}%
\subfigure[rating imputation]{\label{task3_groc}
\includegraphics[width=1.5in]{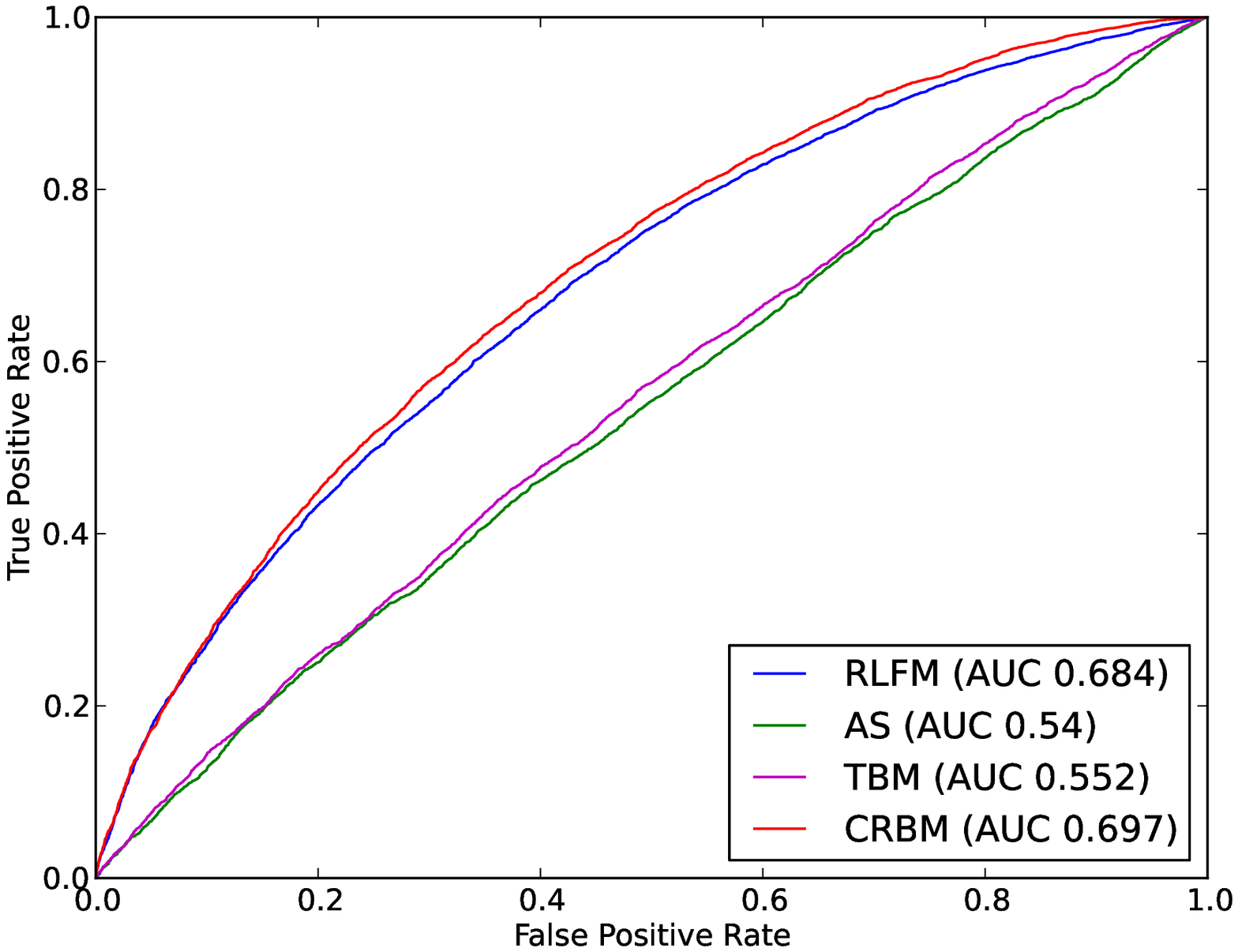}}
\caption{\label{fig_groc}
Performance of different algorithm in three 
recommendation tasks}
\end{figure}

We compare the performance of these four algorithms, including TBM,
 AS, RLFM and CRBM, in three recommendation tasks,
 see figure \ref{fig_groc}.
As the one-class implicit prediction and one-class explicit
 prediction give very similar 
results, we will use one-class prediction for brief. 
According to figure \ref{task1_groc} and figure 
\ref{task2_groc}, we could sort these models by their
performance in descending order and get the result
(TBM, RLFM, AS, CRBM). However, we get
 almost an opposite result when considering 
figure \ref{task3_groc}.
In the following, we discuss how the difference
arise. A key difference between one-class prediction
and rating prediction is how to deal with the missing value.

On one hand, all user-item pairs have a certain value in the
one-class prediction task. 
All missing value between users and items take
value 0 when we train models, see section \ref{recommendation_task}. 
When predicting for cold start items,
all user will be considered. The ROC curve is plotted using all
pairs between users and new items.
On the other hand, we remain the missing state
of the corresponding user-item pair.
When predicting for new items we only test users that have rated them.
The ROC curve is plotted only using user-item pairs in the test set.
Concretely, in one-class prediction there are two states for each
user-item pair 0 and 1, the rating matrix is a dense.
While in rating prediction, the rating matrix is sparse.
This difference is of significance and play an important role for evaluate
the performance of models. In the former case we need to predict
ratings between all user-item pairs.
Problems arise when we fill 0 value to 
the missing pairs which will be illustrated by the following
example. 

Suppose there are 1000 users and 1000 movies in our data set.
Without loss of generality, suppose a user in the test set likes 20 movies.
Now we need evaluate an algorithm using this data set.
In the one-class prediction task
 a perfect model should put these 20 movies on the top of the prediction rank list,
which means we assume this user like these 20 movies better than other movies.
This assumption does not seem rational. It's very likely that the user likes
the other movie better than these 20 ones. 
Putting these 20 ones in the top rank list is not what we really want to do.
We want to know the correct rank of all 1000 movies, but this
is impossible generally as the other 900 movies rating
are missing.
So the problem of rating prediction we caused by its filling missing value.

When comes to the rating prediction, the situation is difference.
We only need predict whether a user will like or not for each of
the test movie which are all given in the test set.
In other words, we just to give a rank list
of the test set but not the whole movie set. 
When we do not know the correct the answer, it's
better to ignore them than to take them as 0.
In the above, we analysis the main problem of rating prediction.
We must realize that we could not ignore the missing in one-class
prediction task just as in rating prediction.
Rating prediction could not replace one-class prediction
completely because the later is suitable 
when there are only posititve feedback and no negative 
feedback. Collaborative filtering techniques, such as use-based CF,
item-based CF or matrix factorization, become invalid in such situation.
These techniques always give some trivial result, all user liking 
all items.

It's easy for us to understand the performance of these four models
as we know about the difference and characteristic about the recommendation
task.
Our data set contains both possitive feedback and negative feedback.
So we should use the rating prediction task to evaluate the performance.
Though AS model and TBM gives perfect resutls in one-class prediction,
they predict ratings in the dense way. They use missing values which
 includes uncertainty.
RFLM and CRBM could deal with the origin sparse raing matrix.
If we also sample some negative feedback from the 
missing pair, they could give good performance in the one-class 
prediction. In our experiment we sample the negative feedback 
about 10 times the possitive sapmle.

In the one-class prediction, we ignore the negative feedback and
use fewer information both in train and test phase. RFLM and CRBM
could deal with this situation. But in the opposite, AS and TBM could
only deal with one-class problem, when we ignore the negative feedback
in the train phase and predict more informative result, they
become invalid, just as the performance of AS and TBM in rating prediction.

%\subsection{Compare RMSE MAE and AUC}
%auc the x-axis is the epach, y-axis is rmse.
\subsection{Interpretation}
In this section, we will show the matrix U we have learned
indeed make sense. 
The matrix U gives us more intuition.
Every row of U could be seen as a feature of movies.
A movie has a feature if the feature's corresponding 
state is active.
Every column could be seen as the correspondence value of
actor or genre. So every movie has a representation
by actors. If two actors are similar (similar means user
has similar rating for movies of these actors), then the
value of the distance of correspond column should be
small. So we could cluster the actors by U.
In this paper we use the k-mean algorithm to cluster
the actors. 
Table \ref{actor_com} illustrates 8 clusters from our
experiment. In each cluster, we show 4 actors or genres.
we could see who and who are more similary with each other
 and in which type of movie one actor is usually appear.
For example, we see Brad Pitt and Kevin Bacon are
more likely appear in the same type movie.
Also, the main movie types \textbf{comedy, crime, war} etc.
are partitioned into different clusters. 

\begin{table*}
\caption{Cluster actors and genres based on U}
\centering
\label{actor_com}
\begin{tabular}{|c|c|} 
\hline 
Topic Number & Actors and Genres\\
\hline 
\multirow{1}{*}{1}  &
\multirow{1}{3.5in}{
Rance Howard,Parker Posey,James Earl Jones,John Cusack
}
\\ \hline
\multirow{1}{*}{2}  &
\multirow{1}{3.5in}{
Children's, Brad Pitt, Kevin Bacon, Western
}
\\ \hline
\multirow{1}{*}{3}  &
\multirow{1}{3.5in}{
Comedy, Paul Herman, Michael Rapaport, John Diehl
}
\\ \hline
\multirow{1}{*}{4}  &
\multirow{1}{3.5in}{
 R. Lee Ermey, Tim Roth, Dermot Mulroney, War
} 
\\ \hline
\multirow{1}{*}{5}  &
\multirow{1}{3.5in}{
 Wallace Shawn, James Gandolfini, Crime, Sci-Fi
}
\\ \hline
\multirow{1}{*}{6}  &
\multirow{1}{3.5in}{
Action, Bruce Willis, Sigourney Weaver, Xander Berkeley
}
\\ \hline
\multirow{1}{*}{7}  &
\multirow{1}{3.5in}{
Adventure, Drama, John Travolta, David Paymer
}
\\ \hline
\multirow{1}{*}{8}  &
\multirow{1}{3.5in}{
Thriller, Paul Calderon, Gene Hackman, Steve Buscemi
}
\\ \hline
\end{tabular}
\end{table*}

\section{Conclusion and Future work}
\label{conclusion}

In this paper, we apply CRBM to solve the new items cold-start
problem. In fact, it could be easily extended for new user
situation. We compare CRBM with other three typical techniques
and show that in the implicit rating prediction and rating
prediction task, CRBM gives comparable performance, while in
the rating imputation task CRBM gives the best result.
According our analysis, the rating imputation task is more
coincidence with the real situation. So CRBM shows its
 superiority to other models.
Such results give us more indication for future researches.
Firstly, RBM-based models are good at extract features, so
we could give more easy explainable result. 
Secondly, this kinds of model are easily to be applied for 
online application. When there are new items, we just need
update the parameters by the reconstruction.
Thirdly, RBM-based model could be easily combined with
deep models to extract more feature for recommendation
\cite{srivastava2013modeling}.

Just as \cite{welling2004exponential}, we do not want to
show CRBM models are more superior than the directed graphical 
models. But the energy-based model give us some more information
in different application.
In the future we will use deep models to solve the cold start problem.
\bibliography{mybib}
\bibliographystyle{plain}
\end{document}